\PassOptionsToPackage{numbers,sort&compress}{natbib}
\documentclass[preprintnumbers,amsmath,amssymb,prd,notitlepage,nofootinbib,superscriptaddress,twocolumn]{revtex4}
\usepackage[utf8]{inputenc}
\usepackage[numbers,sort&compress]{natbib}
\usepackage[normalem]{ulem}
\usepackage{graphicx,xcolor,dcolumn,bm,subfigure,color}
\usepackage[utf8]{inputenc}
\usepackage[hidelinks]{hyperref}
\usepackage{aas_macros}
\usepackage{enumitem}
\usepackage{url}
\usepackage{listings}
\usepackage{fontawesome}

\definecolor{linkcolor}{rgb}{0.,0.3,0.7}
\definecolor{codegreen}{rgb}{0,0.6,0}
\definecolor{codegray}{rgb}{0.5,0.5,0.5}
\definecolor{codepurple}{rgb}{0.58,0,0.82}
\definecolor{backcolour}{rgb}{0.95,0.95,0.92}

\lstdefinestyle{mystyle}{
    commentstyle=\color{codegray},
    keywordstyle=\color{codegreen},
    numberstyle=\tiny\color{gray},
    stringstyle=\color{codepurple},
    basicstyle=\ttfamily\footnotesize,
    breakatwhitespace=false,
    breaklines=true,
    captionpos=b,
    keepspaces=true,
    numbers=none,
    numbersep=5pt,
    showspaces=false,
    showstringspaces=false,
    showtabs=false,
    tabsize=2
}

\hypersetup{breaklinks, colorlinks,
  urlcolor=linkcolor,linkcolor=linkcolor,citecolor=linkcolor}
\newcommand{\codeicon}{{\color{linkcolor}\faFileCodeO}}
\newcommand{\githubmain}{\href{https://github.com/catketchup/class_rot}{{\faGithub}}}
\newcommand{\ini}{\href{https://github.com/catketchup/class_rot/blob/main/explanatory_ROT.ini}{{\codeicon}}}
\newcommand{\notebook}{\href{https://github.com/catketchup/class_rot/blob/main/notebooks_rot}{{\codeicon}}}
\newcommand{\source}[1]{\href{https://github.com/catketchup/class_rot/tree/main/source/#1}{{\codeicon}}}
\newcommand{\python}[1]{\href{https://github.com/catketchup/class_rot/tree/main/python/#1}{{\codeicon}}}

\makeatletter
\renewcommand{\p@subsection}{}
\renewcommand{\p@subsubsection}{}
\makeatother

\begin{document}

\title{Computing Microwave Background Polarization Power Spectra from Cosmic Birefringence}

\author{Hongbo Cai}
\email{hoc34@pitt.edu}
\affiliation{Department of Physics and Astronomy, University of Pittsburgh, Pittsburgh, PA, USA 15260}

\author{Yilun Guan}
\email{yilun.guan@dunlap.utoronto.ca}
\affiliation{Dunlap Institute for Astronomy and Astrophysics, University of Toronto, 50 St. George St., Toronto, ON M5S 3H4, Canada}

\date{\today}

\begin{abstract}
  We present a new publicly available code, \texttt{class\_rot}, which
  modifies \texttt{class} to enable fast non-perturbative calculation
  of cosmic microwave background polarization power spectra due to both isotropic and
  anisotropic polarization rotation from cosmic birefringence. Cosmic birefringence can arise
  from new parity-violating physics  
  such as axion dark matter with a Chern-Simons coupling to photons or
  Faraday rotation due to a primordial magnetic field. 
  Constraints on these effects can be obtained by comparing measurements to 
  precise numerical calculations of the polarization power spectra. 
  We describe the implementation of \texttt{class\_rot} in
  terms of both mathematical formalism and coding architecture. We also provide usage 
  examples and demonstrate the accuracy of the code by comparing with simulations. \githubmain
\end{abstract}
\maketitle
\section{INTRODUCTION} \label{sec:introduction}
Parity-violating physics in the early universe may cause an effect
known as cosmic birefringence, in which photons with different
polarizations travel differently along their propagation paths,
resulting in a net rotation on the polarization directions of cosmic
microwave background (CMB) photons. Such an effect can arise from many
types of beyond-the-Standard-Model physics, such as from the coupling
between axion-like particles and photons through a
Chern-Simons interaction (see, e.g., \cite{Li:2008}), from pseudoscalar
fields introduced in early dark energy models to resolve the Hubble
tension \cite{Capparelli:2020}, or from primordial magnetic fields
through Faraday rotation (see, e.g., \cite{Kosowsky:1996:FR}).

Cosmic birefringence can cause both isotropic and anisotropic
rotation of the microwave background polarization. Since
the polarization field is dominated by an E-mode signal from primordial
density perturbations, small rotations of polarization effectively turn
E-mode into B-mode polarization, leaving observable imprints in the 
polarization power spectra. Isotropic birefringence, in particular, leads to non-zero
parity-odd power spectra in the CMB including TB and EB (see, e.g.,
\cite{Li:2008, Zhai:2020a}). Various experiments have placed
constraints on isotropic rotation angle, such as Planck \cite{Planck:2016soo},
WMAP \cite{2011}, and ACT \cite{ACT:2020frw}. 
The observational challenge in constraining
isotropic birefringence is that its effect is highly degenerate
to that of a calibration error in the orientation of polarized detectors
(see, e.g., \cite{Keating:2013,Kaufman:2014}). 

Anisotropic birefringence, on the other hand, leads
only to parity-even spectra and contributes non-negligibly 
to the B-mode power spectrum. Anisotropic rotation also induces
off-diagonal correlations in the microwave background multipoles, which allows 
reconstruction of the anisotropic rotation field using a quadratic estimator
approach similar to lensing reconstruction of the deflection field (see, e.g.,
\cite{Gluscevic:2009,Yadav:2012a,Namikawa:2017}). Such an effect has been used
to derive observational constraints on anisotropic rotation; for examples, 
Planck \cite{PlanckCollaboration:2016}, BICEP2 / Keck \cite{BICEP2Collaboration:2017}, 
ACT \cite{Namikawa:2020}, and SPT \cite{Bianchini:2020} have all derived upper bounds on 
anisotropic rotation field with a scale-invariant power spectrum.

Despite the physical importance of a possible rotation field, to our knowledge 
no publicly available codes exist that compute CMB power spectra from cosmic
birefringence. Here we present a modified
version of \texttt{class}
\cite{software:class}\footnote{\url{https://github.com/lesgourg/class_public}},
named
\texttt{class\_rot}\footnote{\url{https://github.com/catketchup/class_rot}},
which implements this calculation and allows for fast computation of
the rotated EB, TB, EE, and BB power spectra due to both
isotropic and anisotropic rotation from cosmic birefringence. In particular, we
implement a non-perturbative calculation based on the angular
correlation function of the rotation field \cite{Li:2008,Li:2013}.
Our code has an accuracy better than 1\% at all multipoles from
$l=2$ to $l=4000$, which we verify through comparison with power
spectra of simulated sky maps including random rotation fields. 

This paper is structured as follows. In Sec.~\ref{sec:rotation}, we
describe the basics of cosmic birefringence. In Sec.~\ref{sec:rotated
  ps} we show the non-perturbative calculation method that is implemented in
\texttt{class\_rot}, focusing on the effect of cosmic birefringence on
the CMB power spectra. In Sec.~\ref{sec:code}, we demonstrate the code 
implementation and give usage examples, and we present 
comparisons between the results from \texttt{class\_rot} and numerical
simulations.  
Sec.~\ref{sec:conclusion} provides a brief concluding discussion about the uses
of this code in the context of current and upcoming experiments. 

\section{COSMIC ROTATION FIELD}
\label{sec:rotation}
The rotation effect from cosmic birefringence can be effectively
expressed as a rotation field $\alpha(\hat{\bm{n}})$, which can have
both an isotropic part and an anisotropic part \cite{Zhai:2020a},
given by
\begin{equation}
  \label{eq:alpha}
  \alpha(\hat{\bm{n}})=\bar{\alpha}+\delta \alpha(\hat{\bm{n}}),
\end{equation}
with $\bar{\alpha}$ the isotropic part, and
$\delta \alpha(\hat{\bm{n}})$ the anisotropic part with a zero mean,
\begin{equation}
  \label{eq:rotation parts}
  \expect{\delta \alpha(\hat{\bm{n}})}=0.
\end{equation}
As a result of rotation, Stokes parameter $Q$ and $U$ transform as 
\begin{equation}
  \label{eq:rotation}
  (\tilde{Q} \pm i \tilde{U})(\hat{\bm{n}})=\exp (\pm i 2 \alpha(\hat{\bm{n}}))(Q \pm i U)(\hat{\bm{n}}),
\end{equation}
where we have used tildes to denote rotated quantities.

To illustrate how such a rotation field can arise from parity-violating
physics in the early universe, consider for example a
Chern-Simons-type interaction of photons and axions
with a Lagrangian given by
\begin{equation}
  \label{eq:cs term}
  \mathcal{L}_{c s}=\frac{\beta \phi}{2 M} F^{\mu \nu} \tilde{F}_{\mu \nu},
\end{equation}
where $\beta$ is a dimensionless coupling constant, $\phi$ is the
axion field, $M$ is its mass scale, and $F^{\mu \nu}$ is the
electromagnetic tensor with $\tilde{F}_{\mu \nu}$ being its dual. This term
modifies the Euler-Lagrange equations for electromagnetic field and induces a
rotation in the polarization direction of a photon if $\phi$
varies along its propagation path \cite{1997PhRvD..55.6760C, 1998PhRvD..58k6002C,Leon:2017}, with the rotation
angle given by
\begin{equation}
  \label{eq:alpha and phi}
  \alpha=\frac{\beta}{M} \Delta \phi,
\end{equation}
where $\Delta \phi$ is the change of $\phi$ along the photon path.
In the case that the axion field $\phi$ is spatially
homogeneous, Eq.~\eqref{eq:alpha and phi} introduces an
isotropic rotation field to the CMB; an inhomogeneous axion field
gives an anisotropic rotation field in the CMB.

A convenient way to express an anisotropic rotation field,
$\alpha(\hat{\bm{n}})$, is to expand it in the basis of spherical
harmonics as
\begin{equation}
  \label{eq:alpha alm}
  \delta \alpha(\hat{\bm{n}})=\sum_{L M} \alpha_{L M} Y_{L M}(\hat{\bm{n}}).
\end{equation}
We assume that $\alpha(\hat{\bm{n}})$ follows Gaussian random
statistics, in which case the statistical information of the rotation
field $\alpha(\hat{\bm{n}})$ can be completely specified by its power
spectrum $C_L^{\alpha\alpha}$, given by
\begin{equation}
\label{eq:alpha ps}
  \expect{a_{L M} a_{L' M'}} = \delta_{L L'}\delta_{M M'}C_{L}^{\alpha\alpha}.
\end{equation}
In this paper we only consider a scale-invariant power spectrum of
the anisotropic rotation field, which is physically well-motivated
\cite{2011PhRvD..84d3504C}, though the formalism presented here is broadly
applicable to an arbitrary rotation field power spectrum. Following the convention in \cite{Abazajian:2019eic}, we parametrize a scale-invariant power spectrum as
\begin{equation}
  \label{eq:cl_aa}
  \frac{L(L+1)}{2 \pi} C_{L}^{\alpha \alpha}=A_{C B},
\end{equation}
with $A_{CB}$ the amplitude of the cosmic birefringence power
spectrum\footnote{Note that $A_{CB}$ defined in this paper is $10^{-4}$ times of that in \cite{Namikawa:2020} and $10^{-5}$ of that in \cite{Namikawa:2017}.}.

\section{Impacts on Microwave Background Polarization Power Spectra}
\label{sec:rotated ps}

\begin{figure}[t]
\includegraphics[width=0.45\textwidth]{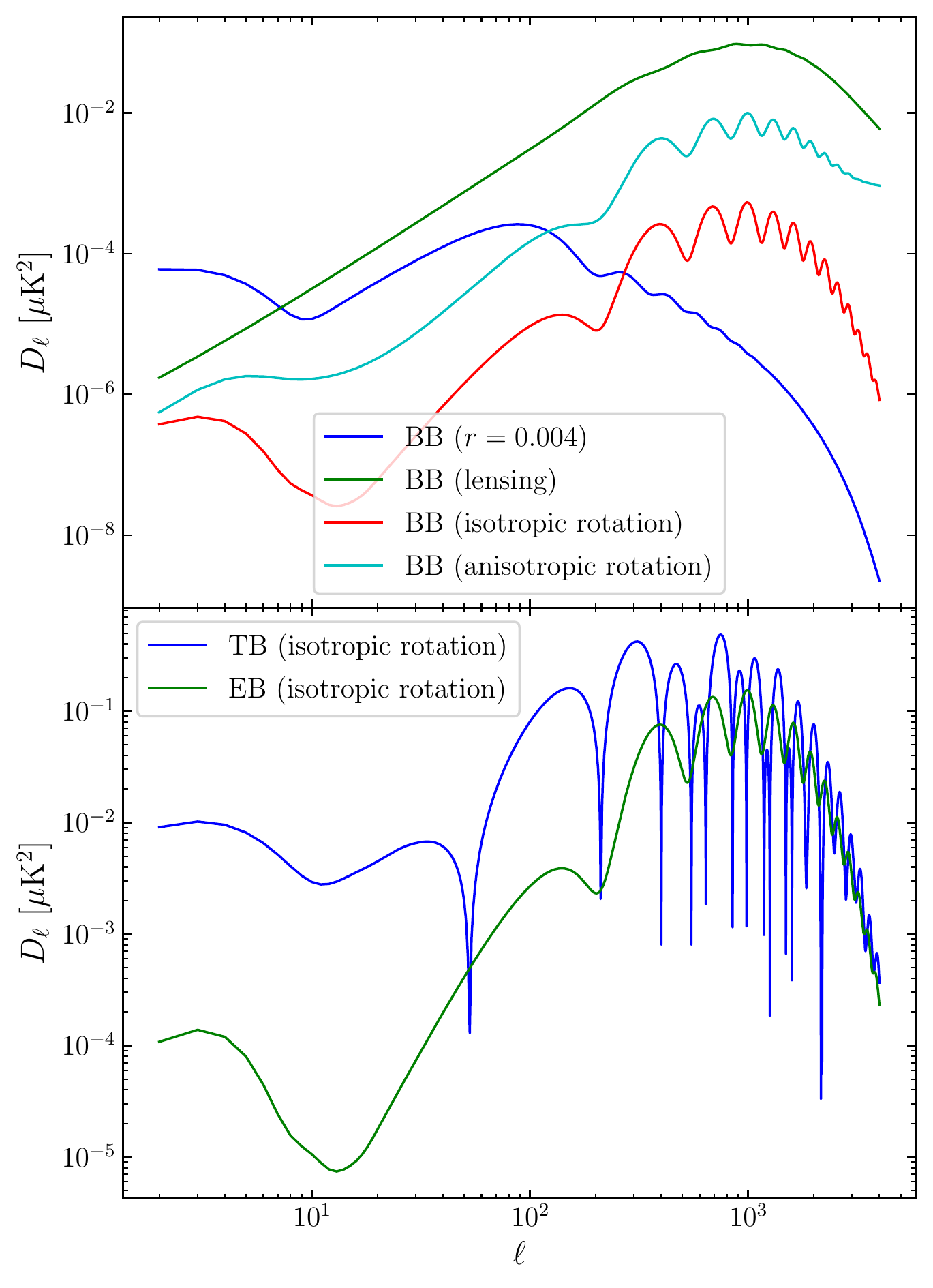}
\caption{Microwave background polarization BB power spectrum contributions from a scale-invariant tensor mode ($r=0.004$), gravitational lensing, isotropic rotation ($\bar{\alpha}=0.1^{\circ}$) and scale-invariant anisotropic rotation ($A_{CB}=10^{-5}$) are given in the upper panel. The absolute TB and EB power spectra from isotropic rotation ($A_{CB}=10^{-5}$) are shown in the lower panel.}
\label{fig:ps.pdf}
\centering
\end{figure}
In this section, we briefly review the rotated CMB power spectra calculation implemented in \texttt{class\_rot}. We consider a rotation field with both an isotropic contribution and an Gaussian random anisotropic contribution as described in Eq.~\eqref{eq:alpha}. We adopt the non-perturbative method introduced in \cite{Li:2008,Li:2013}, which is similar to the calculation method of lensed CMB power spectra in \cite{Challinor:2005}. Here we briefly review the non-perturbative calculations relevant to the implementation of \texttt{class\_rot}; we refer interested readers to \cite{Li:2008,Li:2013} for more calculation details.

In this method, the starting point is to connect the real-space correlation functions of rotated quantities, such as $\tilde{T}(\hat{\bm{n}})$, $\tilde{Q}(\hat{\bm{n}})$, and $\tilde{U}(\hat{\bm{n}})$, to the rotated power spectra, e.g., $\tilde{C}_{\ell'}^{E E}$, $\tilde{C}_{\ell'}^{B B}$, with 
\begin{equation}
  \label{eq:xi spherical}
  \begin{aligned}
    \tilde{\xi}_{+}(\beta) &\equiv\left\langle(\tilde{Q}+i \tilde{U})^{*}(\hat{\bm{n}})(\tilde{Q}+i \tilde{U})\left(\hat{\bm{n}}^{\prime}\right)\right\rangle\\
    &= \sum_{\ell'} \frac{2\ell'+1}{4 \pi}\left(\tilde{C}_{\ell'}^{E E}+\tilde{C}_{\ell'}^{B B}\right) d_{22}^{\ell'}(\beta),\\
    \tilde{\xi}_{-}(\beta) &\equiv\left\langle(\tilde{Q}+i \tilde{U})(\hat{\bm{n}})(\tilde{Q}+i \tilde{U})\left(\hat{\bm{n}}^{\prime}\right)\right\rangle\\
    &= \sum_{\ell'} \frac{2\ell'+1}{4 \pi}\left(\tilde{C}_{\ell'}^{E E}-\tilde{C}_{\ell'}^{B B}+2 i \tilde{C}_{\ell'}^{E B}\right) d_{-22}^{\ell'}(\beta), \\
    \tilde{\xi}_{X}(\beta) &\equiv \left\langle T(\hat{\bm{n}})(\tilde{Q}+i \tilde{U})\left(\hat{\bm{n}}^{\prime}\right)\right\rangle\\
    &= -\sum_{\ell'} \frac{2\ell'+1}{4 \pi}\left(\tilde{C}_{\ell'}^{T E}+i \tilde{C}_{\ell'}^{T B}\right) d_{02}^{\ell'}(\beta),
  \end{aligned}
\end{equation}
where $\hat{\bm{n}}$ and $\hat{\bm{n}}^{\prime}$ are two directions in the spherical coordinate system, $\cos\beta = \hat{\bm{n}} \cdot \hat{\bm{n}}^{\prime}$, and $d_{mm'}^{\ell}$ is the Wigner d-function. Taking advantages of the orthogonality relations of Wigner d-functions,
\begin{equation}
    \label{eq:w-d orthogonality}
    \int_{-1}^{1} d \cos \beta\: d_{mk}^{\ell}(\beta) d_{m'k'}^{\ell'}(\beta) = \frac{2}{2\ell+1}\delta_{mm'}\delta_{kk'}\delta_{\ell \ell'},
\end{equation}
one can invert Eq.~\eqref{eq:xi spherical} to express rotated power spectra in terms of correlation functions, such as
\begin{equation}
  \label{eq:xi reverse}
  \tilde{C}_{\ell}^{E E}+\tilde{C}_{\ell}^{B B}=2 \pi \int_{-1}^{1} d \cos \beta\:\tilde{\xi}_{+}(\beta) d_{22}^{\ell}(\beta) .
\end{equation}
Applying Eq.~\eqref{eq:rotation}, $\tilde{\xi}_{+}(\beta)$ can be expressed by un-rotated quantities as
\begin{equation}
  \label{eq:xi}
    \tilde{\xi}_{+}(\beta) =e^{-4C^{\alpha}(0)+4C^{\alpha}(\beta)}\sum_{\ell'}(2\ell'+1)(C_{\ell'}^{EE}+C_{\ell'}^{BB})d_{22}^{\ell'}(\beta).
\end{equation}
Here $C^{\alpha}(\beta)$ is the correlation function of rotation angles in the two directions separated by $\beta$ and can be expressed as
\begin{equation}
  \label{eq:cla}
  \begin{aligned}
    C^{\alpha}(\beta)=\left\langle\delta \alpha\left(\hat{\bm{n}}_{1}\right) \delta \alpha\left(\hat{\bm{n}}_{2}\right)\right\rangle=&\ \sum_{L} \frac{2 L+1}{4 \pi} C_{L}^{\alpha \alpha} P_{L}(\cos \beta)\\
    =&\ \sum_{L} \frac{2 L+1}{4 \pi} C_{L}^{\alpha \alpha} d_{00}^{L}(\beta),
\end{aligned}
\end{equation}
where $C_{L}^{\alpha \alpha}$ is a generic rotation field power spectrum introduced in Eq.~\eqref{eq:alpha ps}, $P_{L}(\cos \beta)$ is the Legendre Polynomial, and we have applied $P_{L}(\cos \beta) = d_{00}^{L}(\beta)$.

Equipped with Eq.~\eqref{eq:xi}, Eq.~\eqref{eq:xi reverse} can be written as
\begin{equation}
  \label{eq:rotated ps EE BB}
  \begin{aligned}
    \tilde{C}_{\ell}^{E E}+\tilde{C}_{\ell}^{B B} &=\frac{1}{2} e^{-4 C^{\alpha}(0)} \int d\cos \beta\: e^{4C^{\alpha}(\beta)} d_{22}^{\ell}(\beta) \\ &\left[ \sum_{\ell'}(2\ell'+1)(C_{\ell'}^{EE}+C_{\ell'}^{BB})d_{22}^{\ell'}(\beta)\right].
    \end{aligned}
\end{equation}
Similarly, one can also obtain 
\begin{equation}
  \label{eq:rotated ps}
  \begin{aligned}
    \tilde{C}_{\ell}^{T E} &=C_{\ell}^{T E} \cos (2 \bar{\alpha}) e^{-2 C^{\alpha}(0)},\\
    \tilde{C}_{\ell}^{T B} &=C_{\ell}^{T E} \sin (2 \bar{\alpha}) e^{-2 C^{\alpha}(0)},\\
    \tilde{C}_{\ell}^{E E}-\tilde{C}_{\ell}^{B B} &=\frac{1}{2} e^{-4 C^{\alpha}(0)}\cos 4\bar{\alpha}  \int d\cos \beta\: e^{-4C^{\alpha}(\beta)} d_{-22}^{\ell}(\beta)\\ &\left[ \sum_{\ell'}(2\ell'+1)(C_{\ell'}^{EE}-C_{\ell'}^{BB})d_{-22}^{\ell'}(\beta)\right],\\
    \tilde{C}_{\ell}^{E B} &=\frac{1}{2} e^{-4 C^{\alpha}(0)} \sin 4\bar{\alpha} \int d\cos \beta\: e^{-4C^{\alpha}(\beta)} d_{-22}^{\ell}(\beta)\\ &\left[ \sum_{\ell'}(2\ell'+1)(C_{\ell'}^{EE}-C_{\ell'}^{BB})d_{-22}^{\ell'}(\beta)\right].
  \end{aligned}
\end{equation}
Note that the rotated CMB EE, BB and EB power spectra in Eq.~\eqref{eq:rotated ps EE BB} and Eq.~\eqref{eq:rotated ps} are given by real-space integrals, which avoids convolution in the $\ell m$ space which is computationally expensive. A similar strategy that uses real-space integral instead of convolution in $\ell m$ space can be found in delensing calculation \cite{Smith:2012} which significantly reduces computational cost. Also note that we have ignored the correlations between the rotation field and both CMB temperature and (unrotated) E-polarization fields, which may arise in certain axion-like models, such as models with nonzero potential under adiabatic initial conditions \cite{2011PhRvD..84d3504C}. A similar calculation that takes account of these correlations can be found in \cite{Zhai:2020a}.

We can see from Eq.~\eqref{eq:rotated ps EE BB} and Eq.~\eqref{eq:rotated ps} that both isotropic and anisotropic rotations contribute to BB power spectrum. In the upper panel of Fig.~\ref{fig:ps.pdf}, we show the BB power spectrum contributed by an isotropic rotation field with $\bar{\alpha}=0.1^{\circ}$ and a scale-invariant anisotropic rotation field with $A_{CB}=10^{-5}$, respectively. As a comparison, we also show the contributions from primordial tensor mode with $r=0.004$ where $r$ is the tensor-to-scalar ratio, and the contribution from CMB lensing. One can see that the B-mode signal from rotation fields can be larger than that from the primordial tensor mode at $\ell \gtrsim 150$, which suggests that, apart from searching for parity-violating physics, rotation field is also an important systematic when searching for primordial tensor mode. We also note that rotation field generally contributes less than CMB lensing to B-mode polarization; this suggests that the ability to ``de-lens" the CMB will help tighten the constraints on cosmic birefringence.
From Eq.~\eqref{eq:rotated ps} we can also see that both $\tilde{C}_{\ell}^{T B}$ and $\tilde{C}_{\ell}^{E B}$ become non-zero when $\bar{\alpha}$ is non-zero; this is consistent with the fact that an isotropic rotation field violates parity symmetry and induces odd-parity CMB power spectra (see the lower panel of Fig.~\ref{fig:ps.pdf} for example). 

\begin{figure}[t]
\includegraphics[width=0.45\textwidth]{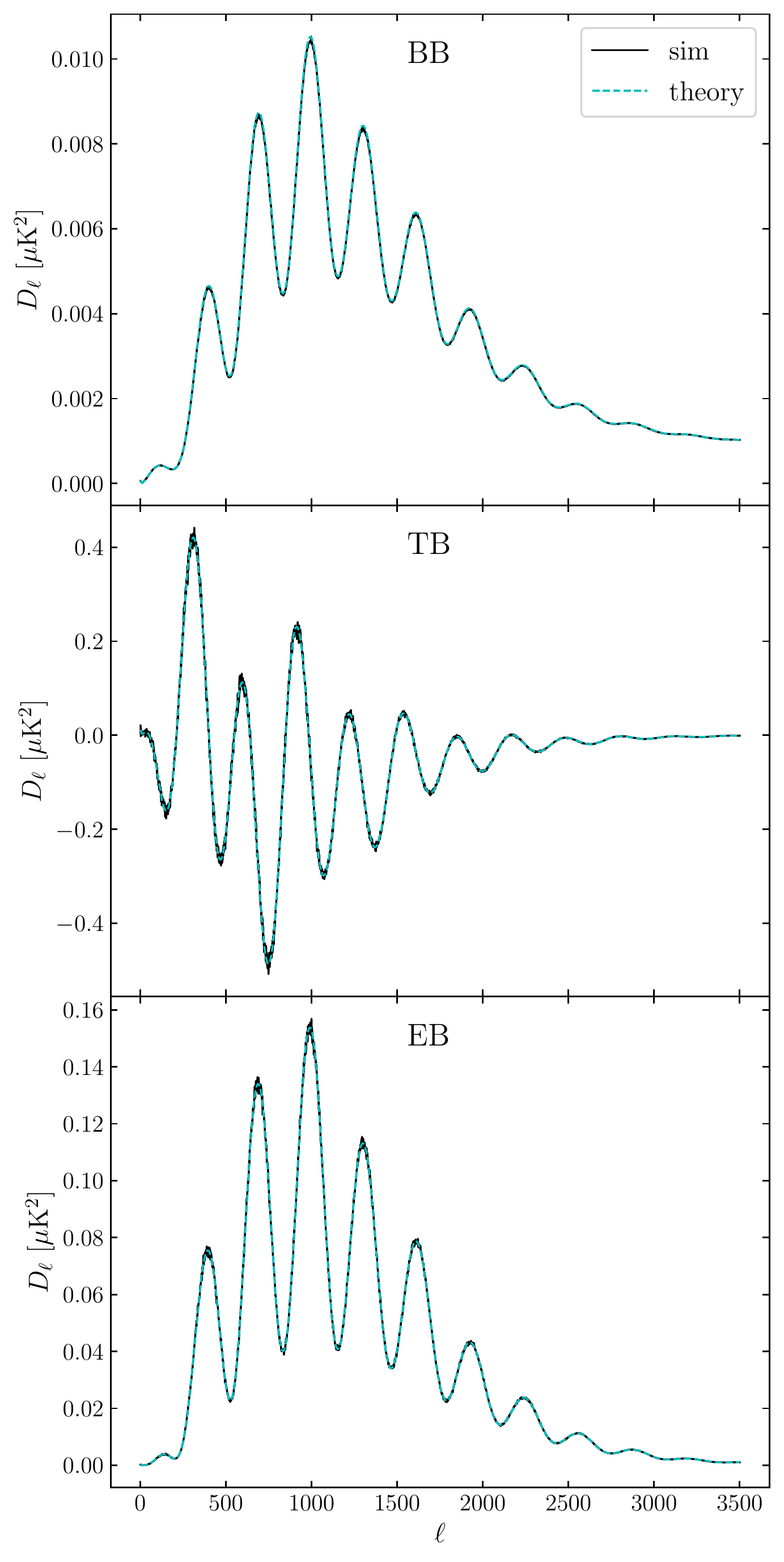}
\caption{rotated CMB BB, TB and EB power spectra from simulation and theory. The theory curves are calculated by \texttt{class\_rot}. The parameters are chosen as: $r=0.004$, $\bar{\alpha}=0.1^{\circ}$ and $A_{CB}=10^{-5}$.}
\label{fig:ps_sims.pdf}
\centering
\end{figure}

\section{The Software Package}
\label{sec:code}
In this section, we describe briefly the implementation of \texttt{class\_rot}, give usage examples of its Python interface, and show comparisons to numerical simulations.
\vspace{0.2cm}

\textbf{Code implementation:}
In \texttt{class\_rot}, the calculations described in Sec.~\ref{sec:rotated ps} are implemented as a new module to \texttt{class}, contained in \texttt{rotation.c} \source{rotation.c}. Internally, this \texttt{rotation} module takes the power spectra calculated from the \texttt{harmonic} module as inputs, by doing so we have implicitly neglected the effect of CMB lensing when calculating the rotated power spectrum. This assumption significantly simplifies our code implementation and will only lead to sub-percent to percent level error due to the smallness of $C_\ell^{BB}$ relative to $C_\ell^{EE}$; to incorporate the effect of CMB lensing in the \texttt{rotation} module will be the subject of future work.

The \texttt{rotation} module can be turned on by specifying \texttt{rotation = yes} in the parameter file, and it can take two additional parameters that specify the rotation field, \texttt{alpha} and \texttt{A\_cb}, which correspond to $\bar{\alpha}$, in unit of degrees, and $A_{CB}$, in radians as defined in Eq.~\eqref{eq:cl_aa}, respectively. The rest of the parameters are identical to those in \texttt{class}. Note that by using $A_{CB}$ we implicitly assume that the rotation field follows a scale-invariant power spectrum -- a choice of preference rather than necessity; other rotation power spectrum can be implemented by changing the \texttt{rotation\_cl\_aa\_at\_l} function defined in \texttt{rotation.c} \source{rotation.c}. We leave the support for taking in a generic rotational power spectrum as input to a future work. 

The parameters can be specified in a parameter file and passed to the compiled \texttt{class} binary executable, in the same way as the original \texttt{class}. An example parameter file, \texttt{explanatory\_ROT.ini} \ini\, is also provided as part of \texttt{class\_rot} to illustrate the use of parameters. Note that this parameter file is only needed when calling \texttt{class\_rot} from the command-line interface using its compiled binary executable. We have also provided Python bindings to the functions in the rotation module allowing them to be called in the Python interface, and we show some usage example below.
\vspace{0.2cm}

\textbf{Usage example:}
Here we give an example of how to calculate the rotated CMB power spectra using the Python interface of \texttt{class\_rot}:
\begin{lstlisting}[language=Python]
from classy import Class

params = {
  "output": "tCl,pCl,rCl",
  "l_max_scalars": 4000,
  "rotation": "yes",
  "alpha": 0.1,
  "A_cb": 1E-5,
}

cosmo = Class()
cosmo.set(params)
cosmo.compute(level=["rotation"])
cosmo.rotated_cl()
\end{lstlisting}
One can see that \texttt{class\_rot} is meant to be used as a drop-in replacement to the original \texttt{class} as it is imported the same way and follows the same usage pattern. The parameters are specified in a Python dictionary, \texttt{param}, and passed to the \texttt{cosmo} object. Note that it is important to include \texttt{rCl} in the \texttt{output} option as it is required for computing the rotated power spectra. The option \texttt{rotation} turns on the rotation module when its value is \texttt{yes}; \texttt{alpha} and \texttt{A\_cb} specify the rotation parameters as can be used in a parameter file. Also note that when computing cosmological model with the function \texttt{cosmo.compute()}, one needs to include \texttt{level=["rotation"]} so that the rotation module and its dependencies are initialized properly. After running \texttt{cosmo.compute()}, the rotated power spectra can be obtained by the function call \texttt{cosmo.rotated\_cl()}, in the form of a Python dictionary following the convention from \texttt{class}. This illustrates a basic usage of \texttt{class\_rot}; we refer interested readers to the examples provided in the bundled Jupyter notebook in \texttt{class\_rot} to find more detailed examples and explanations \notebook.
\vspace{0.2cm}

\textbf{Comparison with simulations:}
To demonstrate the accuracy of \texttt{class\_rot}, we compare the rotated CMB power spectra from \texttt{class\_rot} with those from full-sky simulations. In particular, we first generate 100 realizations of un-rotated CMB maps in T, Q, and U based on a fiducial model given by the best-fit cosmology from Planck 2018 \cite{Planck2018:VI:CP} with $l_{\rm max} = 6000$. Additionally we set a non-zero tensor-to-scalar ratio $r=0.004$. Next we generate 100 realizations of a full-sky rotation map with $\bar{\alpha}=0.1^{\circ}$ and $A_{CB}=10^{-5}$, which are then used to rotate each realization of unrotated CMB maps. These full-sky simulations are generated using \texttt{pixell} \cite{2021ascl.soft02003N} in rectangular pixelization and CAR projection with a resolution of 1 arcminute. We apply each rotation field to rotate one realization of simulated CMB maps in pixel space using Eq.~\eqref{eq:rotation} and then calculate its power spectra after the rotations. We repeat this procedure for each realization to get 100 sets of rotated CMB power spectra.
In Fig.~\ref{fig:ps_sims.pdf}, we show the average of the 100 realizations of rotated power spectra in comparison to the corresponding theory spectrum obtained from \texttt{class\_rot}. One can clearly see that the output of \texttt{class\_rot} is in an excellent agreement with simulations. 
For $C_\ell^{BB}$ we estimate an error of $\lesssim 1\%$ at $\ell\lesssim 4000$; the accuracy noticeably degrades at larger $\ell$ likely due to a combination of pixel effect, numerical precision, and the smallness of the signal of interests. Both $C_\ell^{TE}$ and $C_\ell^{EB}$ from \texttt{class\_rot} agree with the simulations within the expected cosmic variance of the averaged power spectra up to $\ell = 6000$, which is the highest multipole we have tested. 

\section{Discussion and Conclusion}
\label{sec:conclusion}
In this paper we present \texttt{class\_rot}, a new publicly available
modified \texttt{class} code, which calculates rotated CMB power
spectra from cosmic birefringence using a non-perturbative
method. \texttt{class\_rot} supports both isotropic and anisotropic
rotations, as can be specified by the isotropic rotation angle,
$\bar{\alpha}$, and the amplitude of scale-invariant rotation power
spectrum, $A_{CB}$, respectively. Hence, \texttt{class\_rot} can be
effectively used to search for cosmic birefringence signal
that features a scale-invariant rotation power spectrum or an isotropic
rotation in CMB polarization rotation, such as that from the coupling
between axion-like particles and photons via Chern-Simons interaction.
We leave the implementation of a more generic (i.e., not scale-invariant)
rotation power spectrum in \texttt{class\_rot} to a future work which
will allow us to search for a broader range of rotation signal such
as that caused by Faraday rotation from primordial magnetic field, which,
depending on its generation mechanism, may induce a rotation field that is
not scale-invariant (see \cite{2013A&ARv..21...62D} for a review). 

In this paper we have also briefly reviewed the non-perturbative calculation
implemented in \texttt{class\_rot}, which makes use of the angular correlation
function of the rotation field and does not require the rotation to be perturbatively
small. Hence the calculation in \texttt{class\_rot} offers a broader range of
applicability. We leave the implementation of a perturbative calculation as
well as a detailed comparison between the non-perturbative and perturbative methods,
in terms of both speed and accuracy, to a future work.

We have briefly described the coding implementation and given an example of how 
to use \texttt{class\_rot} with its Python interface. To demonstrate its accuracy we 
have compared the rotated CMB power spectra such as BB, TB, and EB obtained 
from \texttt{class\_rot} to full-sky simulations and shown that they are in 
good agreements with $\lesssim 1\%$ error. The upcoming experiments are expected to  
constrain cosmic birefringence with much higher precision. For example, while the current best limits lie around $\mathcal{O}(10')$ for isotropic rotation \cite{Planck:2016soo,ACT:2020frw} and around $\mathcal{O}(10^{-6})$ for $A_{CB}$ \cite{Namikawa:2020,Bianchini:2020}, it has been forecasted that Simons Observatory \cite{SO:2019:SciGoal} can improve the current limits by nearly an order of magnitude, achieving an uncertainty level of around 0.7$'$ for isotropic rotation and around $10^{-7}$ for $A_{CB}$ \cite{Pogosian:2019}. These limits will be further improved by the CMB-S4 experiment \cite{S4:2016:SciBook}, reaching an uncertainty level of around $0.2'$ for isotropic rotation \cite{Abazajian:2019eic} and around $10^{-8}$ for $A_{CB}$ \cite{Pogosian:2019}; this will allow for percent-level determinations of $\bar{\alpha}$ and $A_{CB}$ should there be a cosmic birefringence signal at our current observational limit. In light of these future prospects, it is important to have a robust code that computes the effect of cosmic birefringence in power spectra with better than percent-level accuracy. Hence, \texttt{class\_rot} can be a powerful tool for searches of cosmic birefringence signal in the future.

\section*{Acknowledgments}
We thank Arthur Kosowsky for very helpful comments. We thank Toshiya Namikawa, J. Colin Hill, Mathew S. Madhavacheril, and
Lisong Chen for useful discussion. This work uses resources of the
National Energy Research Scientific Computing Center, and open source
softwares including \texttt{healpy} \cite{2019JOSS....4.1298Z} and
\texttt{pixell} \cite{2021ascl.soft02003N}.

\bibliography{birefringence,lensing,cite}
\end{document}